\documentclass{article}
\usepackage{latexsym}
\newcommand{\be}{\begin{equation}}
\newcommand{\ee}{\end{equation}}
\newcommand{\bear}{\begin{eqnarray}}
\newcommand{\ear}{\end{eqnarray}}

\newcommand{\slpartial}{\raise.15ex\hbox{$/$}\kern-.57em\hbox{$\partial$}}

\def\e{\mbox{e}}
\def\tr{{\rm tr}\,}
\def\slash#1{#1\!\!\!\raise.15ex\hbox {/}}

\def\bea{\begin{eqnarray}}\def\barr{\begin{array}}\def\earr{\end{array}}
\def\eea{\end{eqnarray}}

\begin{document}
\setlength{\unitlength}{1mm}

\begin{titlepage}

\begin{flushright}
Edinburgh 2001/07\\
LAPTH--853/01\\
LPT--Orsay 01--55\\
UMSNH--PHYS/01--7\\
June 2001
\end{flushright}
\vspace{1.cm}

\begin{center}
\large\bf
{\LARGE\bf Calculation of 1--loop Hexagon Amplitudes  \\
in the Yukawa Model}\\[1cm]
\rm
{T.~Binoth$^{a}$, J.~Ph.~Guillet$^{b}$, 
 G.~Heinrich$^{c}$ and C.~Schubert$^{d}$}\\[1cm]

{\em $^{a}$Department of Physics and Astronomy,\\
           The University of Edinburgh,
	   EH9 3JZ Edinburgh, Scotland} \\[.5cm]

{\em $^{b}$Laboratoire d'Annecy-Le-Vieux de Physique 
 Th\'eorique\footnote{UMR 5108 du CNRS, associ\'ee \`a 
              l'Universit\'e de Savoie.} LAPTH,}\\
      {\em Chemin de Bellevue, B.P. 110, F-74941, 
           Annecy-le-Vieux, France} \\[.5cm]   
	    
{\em $^{c}$Laboratoire de Physique Th\'eorique\footnote{UMR 8627 du CNRS.} LPT\\
           Universit\'e de Paris XI, B\^atiment 210, 91405 Orsay, France} \\[.5cm]
    
{\em $^{d}$Instituto de Fisica y Matem\'aticas\\
           Universidad Michoacana de San Nicol\'as de Hidalgo\\
	   Apdo. Postal 2--82\\
	   C.P. 58040, Morelia, Michoac\'an, M\'exico} \\[.5cm]

\end{center}
\normalsize

%

\vspace{1cm}

\begin{abstract}
We calculate a class of one--loop six--point amplitudes 
in the Yukawa model. The construction of multi--particle
amplitudes is done in the string inspired formalism 
and compared to the Feynman diagrammatic approach. 
We show that there exists a surprisingly efficient 
way of calculating such amplitudes 
by using cyclic identities of kinematic coefficients and discuss in detail
cancellation mechanisms of spurious terms.   
A collection of formulas which are useful for the calculation of
massless hexagon amplitudes is given. 
\end{abstract}



\end{titlepage}

\section{Introduction}

Present and future collider experiments will provide
us with more and more experimental data containing
information on multi--particle final states. 
With increasing precision the quest for the inclusion
of quantum corrections arises. This is especially true
for QCD observables, as calculations on the 
Born level are typically plagued by large scale uncertainties  
and therefore  are hardly predictive.

Whereas next-to-leading order calculations for $2\rightarrow 3$
processes have become available in the last years \cite{BernDixonKosower,KunsztTrocsanyi,
BernDixonKosowerWeinzierl_amps,CampbellGloverMiller_amps}, the step to
$2\rightarrow 4$ processes, or even higher, has not been made yet.
The reason lies in the fact that the computation of the 
corresponding amplitudes is highly nontrivial. 
Although iterative reduction methods allowing for a brute force approach
are understood for such amplitudes \cite{Npointpaper}, it turns out
that it is necessary to understand
better recombination and simplification mechanisms
at intermediate steps of the calculation in order to avoid intractably large
expressions.

In the last few years methods either
directly based on string perturbation theory
\cite{berkos,bedush} 
or on a string-like rewriting of field theory amplitudes
\cite{strassler} have been used to derive a number of
``master formulas'' for one-loop $N$--point amplitudes.
Those are generating functionals which upon expansion
yield, for any $N$, 
a closed parameter integral expression for the amplitude.
At the one-loop level, master formulas have
been derived for the 
QCD gluon amplitudes on--shell
\cite{berkos} as well as off--shell \cite{frimagrus}, 
the scalar/spinor QED photon amplitudes 
in vacuum \cite{berkos,strassler} as well as in a constant field
\cite{shaisultanov}, 
the graviton amplitudes in quantum gravity \cite{bedush},
and for amplitudes involving a fermion loop and either
vectors and axialvectors \cite{dimcsc} 
or scalars and pseudoscalars \cite{review}.
A multi-loop generalisation exists for the case 
of the QED photon amplitudes \cite{ss3}.
The resulting integral representations are related
to standard Feynman parameter integrals in a well-understood 
way \cite{berdun}. Nevertheless, 
due to their superior organisation they
often allow one to exploit at the integral
level properties of an amplitude which normally would be seen only
at later stages in a Feynman graph calculation \cite{review}.

Although the string inspired formalism allows for an elegant
formulation of amplitudes in terms of  a manifest Lorentz
structure one is, except in certain particularly 
favourable cases \cite{bedush,ss3}, not at all dispensed from 
doing cumbersome algebraic work. 
The complexity of doing tensor reduction in momentum space
translates into the need to reduce Feynman parameter integrals with
nontrivial numerators to genuine $N$--point scalar
integrals. Finally these have to be reduced further down to 
known scalar integrals.  
Substantial cancellations are typical in all these steps
and progress in finding efficient calculation methods
relies on a better understanding of these mechanisms.

The calculation presented here was triggered by the
observation that the string inspired master
formulas derived in \cite{review} for the one-loop
$N$--point functions in the Yukawa model allow
one to directly express these Green's functions in
terms of scalar parameter integrals, without
the appearance of tensor integrals which one would normally 
expect to encounter in a parameter integral computation of a fermion loop
amplitude. Their computation should therefore be considerably
simpler than the one of the $N$--photon or gluon amplitudes,
so that their study can serve as an intermediate step towards the 
computation of gauge theory amplitudes. 

In section 2 we will present the construction of the $N$--point 
scalar amplitude coming from the Yukawa couplings in the
string inspired formalism. In section 3 we will formulate
the same amplitude in terms of Feynman diagrams
and show the equivalence of the expressions. Then we
will derive a compact expression for the amplitude 
with special emphasis on the cancellation mechanisms 
at work. Section 4 contains our conclusions.

\section{Constructing multi--particle amplitudes 
in the string inspired formalism}

The minimal setting for the amplitudes in question is
a Yukawa Model with both 
a scalar $\phi$ and a pseudoscalar $\phi_5$,
\be
{\cal L}_{\rm yuk}
=
\bar\psi
\Bigl[
i\slpartial
-m 
-g\phi
-ig_5\gamma^5\phi_5
\Bigr]
\psi
+ {\cal L}_{\rm \phi,\phi_5}
\label{defyukawamodel}
\ee
We did not write out the scalar/pseudoscalar part
of the Lagrangian, ${\cal L}_{\rm \phi,\phi_5}$, 
as it is not used in the following.
Still we want to note that the quartic interactions have to be included if the
model is to be renormalizable, since the (pseudo)scalar
four-point functions are divergent.

Based on earlier work on the worldline representation of Yukawa couplings
(refs. \cite{mnss,dhogag}), in \cite{review} the following master formulas
for the fermion loop contributions to the one-particle
irreducible one-loop amplitudes with an {\sl even} number of legs $N$ 
have been derived:
For the $N$ -- scalar case this formula reads\footnote{Note that in contrast to \cite{review} we use here the metric signature $(+,-,-,-)$)},
\bear
\Gamma_{\rm yuk}^{\phi}[p_1,\ldots,p_N]
&=&
-2
(i\,g)^N
{\displaystyle\int_{0}^{\infty}}{dT\over T}
\e^{-m^2T}
{(4\pi T)}^{-{n\over 2}}
\prod_{i=1}^N \int_0^T 
d\tau_i
\int
d\theta_1\cdots\int d\theta_N
\int d\varepsilon_N\cdots\int d\varepsilon_1
\nonumber\\
&&
\hspace{-40pt}
\times\exp\biggl\lbrace
-
\sum_{i,j=1}^N
\Bigl\lbrack
{1\over 2}
(G_{Bij}+\theta_i\theta_j G_{Fij})
p_i\cdot p_j
+{1\over 2}
(G_{Fij} + \theta_i\theta_j
2\delta_{ij}
)
\varepsilon_i\varepsilon_j\Bigr]
+2\,i\,m\sum_{j=1}^N
\varepsilon_j\theta_j
\biggr\rbrace
\nonumber\\
\label{yukawascalarmaster}
\ear
Here $T$ is the global proper-time variable for
the loop fermion, and
$G_{Bij},G_{Fij},\delta_{ij}$ denote the
basic worldline Green's functions
\cite{review}
\bear
G_{Bij}&=&
\mid \tau_i-\tau_j\mid 
-{{(\tau_i-\tau_j)}^2\over T}
\nonumber\\
G_{Fij}&=&
{\rm sign}(\tau_i-\tau_j)
\nonumber\\
\delta_{ij} &=&
\delta(\tau_i-\tau_j)
\nonumber\\
\label{defgreens}
\ear
The Grassmann variables $\theta_1,\ldots,\theta_N$
and ``polarisation scalars''
\footnote{Those appear here because the derivation
of these formulas uses dimensional
reduction from six-dimensional gauge theory. For details see
\cite{review,mnss}.} 
$\varepsilon_1,\ldots,\varepsilon_N$
are all anticommuting with each other, as well as with
$d\theta_1,\ldots,d\theta_N$
and
$d\varepsilon_1,\ldots,d\varepsilon_N$.
The Grassmann integration rules are 
$\int d\theta_i \theta_i = \int d\varepsilon_i \varepsilon_i =1$.

The corresponding formula for the pseudoscalar
case is somewhat simpler,
\bear
\Gamma_{\rm yuk}^{\phi_5}[p_1,\ldots,p_N]
&=&
-2
(i\,g_5)^N
{\displaystyle\int_{0}^{\infty}}{dT\over T}
\e^{-m^2T}
{(4\pi T)}^{-{n\over 2}}
\prod_{i=1}^N \int_0^T 
d\tau_i
\int
d\theta_1
\cdots \int d\theta_N
\int d\varepsilon_N\cdots\int d\varepsilon_1
\nonumber\\
&&
\hspace{-25pt}
\times\exp\biggl\lbrace
-
\sum_{i,j=1}^N
\Bigl\lbrack
{1\over 2}
(G_{Bij}+\theta_i\theta_j G_{Fij})
p_i\cdot p_j
+{1\over 2}
(G_{Fij} + \theta_i\theta_j
2\delta_{ij}
)
\varepsilon_i\varepsilon_j\Bigr]
\biggr\rbrace
\nonumber\\
\label{yukawapseudoscalarmaster}
\ear
Note that these formulas are also valid off-shell.
In the massless case, which is the one which we
are going to exploit in the present paper, 
both formulas coincide, as they should
since here the chiral symmetry is unbroken.
In contrast to, e.g., the master formula for the $N$ -- photon
amplitude, for the case at hand the
result of the Grassmann integrations is easy to write down in
closed form. It is a matter of simple combinatorics
to see that
\bear
\int d\theta_1 \cdots d\theta_N
\int d\varepsilon_N \cdots  d\varepsilon_1
\exp\biggl\lbrace
-
\sum_{i,j=1}^N
\Bigl\lbrack
{1\over 2}
(G_{Bij}+\theta_i\theta_j G_{Fij})
p_i\cdot p_j
+{1\over 2}
(G_{Fij} + \theta_i\theta_j
2\delta_{ij}
)
\varepsilon_i\varepsilon_j\Bigr]
\biggr\rbrace
\nonumber\\ 
=
\sum_{j=0}^{N/2}\,Y_{N,N-j}
\exp \Bigl[-\sum_{j<l=1}^N G_{Bjl}p_j\cdot p_l\Bigr]
\label{expandexplicit}
\ear
where
\bear
Y_{N,N-j} &=&
{(-1)^{N/2-j}\over j!(N-2j)!}
\sum_{\pi\in S_{N}}
\delta_{\pi_1\pi_2}
\delta_{\pi_3\pi_4}
\cdots
\delta_{\pi_{2j-1}\pi_{2j}}
{\rm Alt}(\sigma_{\pi_{2j+1}\pi_{2j+2}},\sigma_{\pi_{2j+3}\pi_{2j+4}},
\ldots,\sigma_{\pi_{N-1}\pi_{N}})\nonumber\\&&\hspace{90pt}\times
{\rm Alt}(\slash\sigma_{\pi_{2j+1}\pi_{2j+2}},
\slash\sigma_{\pi_{2j+3}\pi_{2j+4}},
\ldots,\slash\sigma_{\pi_{N-1}\pi_{N}})
\label{YN}
\ear
Here we used the abbreviations
\bear
\sigma_{ij} &\equiv& {\rm sign}(\tau_i -\tau_j) \nonumber\\
\slash\sigma_{ij} &\equiv& \sigma_{ij} \; p_i\cdot p_j \nonumber\\
{\rm Alt}(T_{j_1\,j_2},T_{j_3\,j_4},
\cdots,T_{j_{2k-1}\,j_{2k}})
&\equiv&
{1\over k!2^k}
\sum_{\pi\in S_{2k}}
{\rm sign} (\pi)
T_{\pi_{j_1}\pi_{j_2}}
T_{\pi_{j_3}\pi_{j_4}}
\cdots
T_{\pi_{j_{2k-1}}\pi_{j_{2k}}}
\nonumber\\
\label{abbreviations}
\ear
Note that the variable $j$ counts the numbers of ``pinches''
in a term, so that $(N-j)$ is the number of nontrivial integrations.
This integral represents the whole amplitude; the usual
summation over ``crossed'' diagrams here is replaced by the
integration over the various ordered sectors of the
$N$ -- fold integral. For a fixed ordering any given term
in $Y_{N,N-j}$ produces just a factor times the
standard $(N-j)$--point scalar integral
$I_{N-j}^n$ in $n$ dimensions. 
For the standard ordering $\tau_1>\cdots > \tau_N$ one has
\bear
{\rm Alt}(\sigma_{\pi_1\pi_2}\cdots \sigma_{\pi_{n-1}\pi_{n}})
&=&
{\rm sign}(\pi) \label{idsigmaempty}\\
{\rm Alt}(\slash\sigma_{\pi_1\pi_2}\cdots 
\slash\sigma_{\pi_{n-1}\pi_{n}})
&=&
{\rm sign}(\pi) 
{1\over 4}{\rm tr}(p_1,\dots, p_n)
\label{idsigmafull}
\ear
Inside the traces contraction of the momenta with Dirac matrices
is understood throughout this paper. 
Thus in the massless case we
can write the contribution from this standard sector as
\bear
\Gamma_{\rm yuk}^{1\ldots N}[p_1,\ldots,p_N]=
-{g_{(5)}^N\over (4\pi)^{n/2}}\frac{1}{N}{\cal A}(p_1,\ldots,p_N)=
-{g_{(5)}^N\over (4\pi)^{n/2}}\frac{1}{N}\sum_{j=0}^{N/2}
{\cal A}_{N,N-j}(p_1,\ldots,p_N)
\hspace{1cm}\nonumber\\
2{\cal A}_{N,N-j}(p_1,\ldots,p_N)=
{(-1)^j N\over j!(N-2j)!}
\sum_{\pi\in S_N}
{\rm tr}(p_{\pi_{2j+1}},\dots, p_{\pi_N})
\int_0^{\infty}{dT\over T^{n/2+1}}
\hspace{2cm}\nonumber\\ \times
\int_0^Td\tau_1\int_0^{\tau_1}d\tau_2\cdots \int_0^{\tau_{N-1}}
d\tau_N\, \delta_{\pi_1\pi_2}\cdots\delta_{\pi_{2j-1}\pi_{2j}}
\exp \Bigl[-\sum_{j<l=1}^N G_{Bjl}p_j\cdot p_l\Bigr]
\nonumber\\
=\sum_{j\,\, {\rm pairs}}{\rm tr}(p_1,\dots,
p_{r_1-1}, 
p_{r_1+2},
\dots ,
p_{r_j-1},p_{r_j+2},
\dots,
p_N)
\hspace{3cm}\nonumber\\
\times
I_{N-j}^n(p_1,\ldots,p_{r_1-1},p_{r_1}+p_{r_1+1},
p_{r_1+2},\ldots,
p_{r_j}+p_{r_j+1},\ldots,p_N)
\nonumber\\
\label{finstandard}
\ear
Here in the last expression the sum runs over
all possible ways of deleting $j$ pairs of
adjacent $p_j$'s from the trace
$\tr (p_1,\dots, p_N)$ (including the pair $p_N,p_1$).
Note that in the standard sector only $\delta$--functions with 
adjacent indices contribute,
and for those we have to take a factor of $1/2$
into account since each contribution is shared between
two adjacent sectors.

The whole amplitude is obtained by summing
over all permutations of 
$\Gamma_{\rm yuk}^{1\ldots N}[p_1,\ldots,p_N]$
in $p_1,\ldots,p_N$. Due to cyclic and parity
invariance this sum contains only
$(N-1)!/2$ different terms.

Thus the master formula allows us to directly express
this $N$--point amplitude in terms of scalar integrals.
Note that from the master formulas it is clear that, in the
pseudoscalar case, we can generalise this result to the massive
case simply by replacing all the scalar integrals by their massive
counterparts. In the scalar case, to the contrary, a number of
additional integrals would appear. 

In the following we will focus on the massless case with $N=6$.
All amplitudes with six scalars/pseudoscalars are related by chiral 
invariance. Setting $g=g_5$ one has    
\bear
\Gamma_{\rm yuk}^{\phi}[p_1,p_2,p_3,p_4,p_5,p_6] =
  \Gamma_{\rm yuk}^{6\phi} = \Gamma_{\rm yuk}^{4\phi,2\phi_5} 
= \Gamma_{\rm yuk}^{2\phi,4\phi_5} = \Gamma_{\rm yuk}^{6\phi_5}
\ear
Hence it is enough to compute one of these amplitudes only.

\section{Calculation of the hexagon amplitude $\Gamma_{\rm yuk}^{\phi}$}
Now we turn to the calculation of the  hexagon amplitude 
$\Gamma_{\rm yuk}^{\phi}[p_1,p_2,p_3,p_4,p_5,p_6]$. 
First we rederive the amplitude in the 
Feynman diagrammatic approach. 
The amplitude can be written as a sum over $6!$ permutations
of the external momentum vectors $p_1,\dots,p_6$
\begin{equation}\label{Eq:start}
\Gamma^{\phi}_{\rm yuk}[p_1,p_2,p_3,p_4,p_5,p_6] = -\frac{g^6}{(4\pi)^{n/2}}
\frac{1}{6} 
 \sum\limits_{\pi\in S_6} {\cal A}(p_{\pi_1},p_{\pi_2},
                  p_{\pi_3},p_{\pi_4},p_{\pi_5},p_{\pi_6})
\end{equation} 
Each permutation corresponds to a single Feynman diagram. The amplitude 
for the trivial permutation is given by 
\begin{eqnarray}\label{amp_graph}
{\cal A}(p_1,p_2,p_3,p_4,p_5,p_6) &=& \int \frac{d^nk}{i \pi^{n/2}}
\frac{{\rm tr}(q_1,q_2,q_3,q_4,q_5,q_6)}{q_1^2q_2^2q_3^2q_4^2q_5^2q_6^2}
\end{eqnarray}
where $q_j = k - r_j = k-p_1-\dots-p_j $.
The trivial permutation corresponds to the standard ordering 
of the worldline integral (\ref{finstandard}).
From translation invariance it follows that
\begin{eqnarray}
{\cal A}(p_j,p_{j+1},p_{j+2},p_{j+3},p_{j+4},p_{j+5}) =
{\cal A}(p_{j+1},p_{j+2},p_{j+3},p_{j+4},p_{j+5},p_{j+6})
\end{eqnarray}
Hence it is enough to sum in (\ref{Eq:start}) only over  
non cyclic permutations, i.e. $\pi\in S_6/Z_6$, and to multiply
by a factor 6.
Note that all indices labelling momenta are
understood to be mod 6 throughout this paper. 
Working out the trace gives a sum of products of terms 
$q_k\cdot q_j$ which can be written as $(j>k)$
\begin{equation}
2 q_k\cdot q_j = -(q_k-q_j)^2 + q_k^2 + q_j^2 
               = -(p_j+p_{j-1}+\dots+p_{k+1})^2 + q_k^2 + q_j^2 
\end{equation}
It is now immediately clear that the whole loop momentum dependence
in the numerators is only through inverse propagators which cancel
directly. This means that each graph can simply be represented as
a linear combination of scalar integrals. 
For the trivial permutation we find
\begin{eqnarray}\label{amp_si}
2 \,{\cal A}(p_1,p_2,p_3,p_4,p_5,p_6) =
  {\rm tr}(1) \, I_3^n(p_{12},p_{34},p_{56}) 
 +{\rm tr}(1)\, I_3^n(p_{23},p_{45},p_{61})  \nonumber\\
 +{\rm tr}(p_1,p_2)\, I_4^n(p_1,p_2,p_{34},p_{56}) 
 +{\rm tr}(p_2,p_3)\, I_4^n(p_2,p_3,p_{45},p_{61}) \nonumber\\
 +{\rm tr}(p_3,p_4)\, I_4^n(p_3,p_4,p_{56},p_{12})  
 +{\rm tr}(p_4,p_5)\, I_4^n(p_4,p_5,p_{61},p_{23}) \nonumber\\
 +{\rm tr}(p_5,p_6)\, I_4^n(p_5,p_6,p_{12},p_{34}) 
 +{\rm tr}(p_6,p_1)\, I_4^n(p_6,p_1,p_{23},p_{45})  \nonumber\\
 +{\rm tr}(p_1,p_4)\, I_4^n(p_1,p_{23},p_4,p_{56}) 
 +{\rm tr}(p_2,p_5)\, I_4^n(p_2,p_{34},p_5,p_{61}) \nonumber\\
 +{\rm tr}(p_3,p_6)\, I_4^n(p_3,p_{45},p_6,p_{12})  \nonumber\\
 +{\rm tr}(p_1,p_2,p_3,p_4)\, I_5^n(p_{56},p_1,p_2,p_3,p_4)  
 +{\rm tr}(p_2,p_3,p_4,p_5)\, I_5^n(p_{61},p_2,p_3,p_4,p_5)  \nonumber\\
 +{\rm tr}(p_3,p_4,p_5,p_6)\, I_5^n(p_{12},p_3,p_4,p_5,p_6)  
 +{\rm tr}(p_4,p_5,p_6,p_1)\, I_5^n(p_{23},p_4,p_5,p_6,p_1)  \nonumber\\
 +{\rm tr}(p_5,p_6,p_1,p_2)\, I_5^n(p_{34},p_5,p_6,p_1,p_2)  
 +{\rm tr}(p_6,p_1,p_2,p_3)\, I_5^n(p_{45},p_6,p_1,p_2,p_3)  \nonumber\\
 +{\rm tr}(p_1,p_2,p_3,p_4,p_5,p_6)\, I_6^n(p_1,p_2,p_3,p_4,p_5,p_6) 
\end{eqnarray}
in agreement with formula (\ref{finstandard}) derived in the 
string inspired formalism. 
The arguments of the $N$--point scalar integrals are the momenta
of the external legs. We use the abbreviation 
$p_{ijk\dots}=p_i+p_j+p_k+\dots$.  
The spinor traces can be expressed by Mandelstam variables 
defined by the 9 cuts of the hexagon graph, but the form given above
is not only most compact but also most convenient to proceed.
\begin{eqnarray}
{\rm tr}(1) &=& 4\nonumber\\
{\rm tr}(p_i,p_j) &=&  2 s_{ij} \nonumber\\
{\rm tr}(p_1,p_4) &=& 2 s_{14} = 2 (s_{23} + s_{56} - s_{123} - s_{234}) \nonumber\\
{\rm tr}(p_1,p_2,p_3,p_4) &=& s_{12}(s_{234}-s_{23}) + s_{23}(s_{56}-s_{34})
 + s_{123}(s_{34}-s_{234})\nonumber\\
{\rm tr}(p_1,p_2,p_3,p_4,p_5,p_6) &=&  s_{123} s_{234} s_{345}
      - s_{12} s_{45} s_{234} - s_{23} s_{56} s_{345} - s_{34} s_{123} s_{61}
\end{eqnarray}
with $s_{l,l+1} = ( p_l + p_{l+1})^2$, 
$s_{l-1,l,l+1} = ( p_{l-1} + p_l + p_{l+1})^2$. 
The remaining traces are defined by cyclic relabelling.
In the following the momenta inside the traces will be represented
by their indices only.

Before turning to the computation of the amplitude, we first note
that the amplitude is free of infrared poles. This can be seen
by power counting for the soft and collinear poles separately. 
To investigate the soft limit we 
 replace  in (\ref{amp_graph}) $k^\mu$ by
$\lambda k^\mu$ and let $\lambda \rightarrow 0$. As the integrand times the
measure behave as $\lambda^4d\lambda/\lambda^4$, 
no poles related with the soft limit $\lambda \rightarrow 0$ are present.
To see if collinear limits lead to a divergence it is enough
to study the limit $k||p_1$, i.e. $(k- p_1)^2 \rightarrow 0$
with $|k^\mu|$ nonzero for at least one component. 
To do so we parametrise the loop momentum as
\begin{equation}
k^\mu = z p_1^\mu + \frac{k^2 + k_T^2}{2 \,z \,n\cdot p_1} n^\mu + k_T^\mu 
\end{equation} 
Here $n^\mu$ is an arbitrary light-like four vector not collinear to $p_1$
with\footnote{By Lorentz invariance one can choose
$p_1 = p_1^0 (1,\vec 0^T,1)$.} $\vec n_T=0$. 
The only dangerous propagator in this collinear limit 
is $(k-p_1)^2=-(k^2(1-x)+k_T^2)/x$. It is
easy to see that the numerator is proportional to $k_T$ 
and thus the $k_T$ integration
does not lead to a pole in the collinear limit 
$k\rightarrow x p_1$, since the integral behaves as
\begin{eqnarray}
\int\limits_0^{} dk_T^2 ( k_T^2 )^{-1/2+\epsilon} ( \mbox{const.} + {\cal O}(k_T^2)  )
\end{eqnarray}
Physically speaking the collinear splitting of a massless
spin 0 particle into a massless fermion anti-fermion pair
is infrared safe. 

We turn now to the explicit calculation of the hexagon amplitude. 
We will draw special attention to the cancellation
mechanisms of the spurious poles and to
spurious finite terms. First we
will reduce hexagon and pentagon integrals to box integrals.
Then the explicit expressions for the box integrals are inserted. 
Finally the coefficients of  different terms are combined and simplified
by using linear relations for the reduction coefficients. We note
already that in none of these steps the size of the expression 
will blow up.

To deal with the $N$--point scalar integrals one has to use reduction
formulas \cite{BernDixonKosower,Npointpaper}. Pentagon integrals
can always be represented in  terms of box integrals plus a term
which is of order $\epsilon$, while 
hexagon integrals decay 
into pentagon integrals. Following \cite{Npointpaper},
the reduction formula for the hexagon integral reads:

\newpage

\begin{eqnarray}\label{hexa_red}
I_6^n(p_1,p_2,p_3,p_4,p_5,p_6) = \sum\limits_{j=1}^{6} b_j 
I_5^n(p_{j}+p_{j+1},p_{j+2},p_{j+3},p_{j+4},p_{j+5}) \hspace{3cm} \\
= \sum\limits_{j=1}^{6} \frac{1}{\det(\hat S)}\Big[ 
 {\rm tr}(123456)\,{\rm tr}(\,j+2,j+3,j+4,j+5\,) \hspace{2cm}\nonumber\\
 - 2 s_{j+2,j+3}\,s_{j+3,j+4}\,s_{j+4,j+5}\, 
 {\rm tr}(\,j+5,j,j+1,j+2\,)\Big]
  I_5^n(p_{j}+p_{j+1},p_{j+2},p_{j+3},p_{j+4},p_{j+5})  \nonumber\\
 \nonumber\\
=\frac{1}{\det(\hat S)}
\bigl\{
\bigl[ {\rm tr}(123456){\rm tr}(3456)-2 s_{34}s_{45}s_{56} {\rm tr}(6123) \bigr]\; I_5^n(p_{12},p_3,p_4,p_5,p_6)\nonumber \\
+\bigl[{\rm tr}(123456){\rm tr}(4561)-2 s_{45}s_{56}s_{61} {\rm tr}(1234) \bigr]\; I_5^n(p_{23},p_4,p_5,p_6,p_1)\nonumber\\
+\bigl[{\rm tr}(123456){\rm tr}(5612)-2 s_{56}s_{61}s_{12} {\rm tr}(2345) \bigr] \;I_5^n(p_{34},p_5,p_6,p_1,p_2)\nonumber\\
+\bigl[{\rm tr}(123456){\rm tr}(6123)-2 s_{61}s_{12}s_{23} {\rm tr}(3456) \bigr]\; I_5^n(p_{45},p_6,p_1,p_2,p_3)\nonumber\\
+\bigl[{\rm tr}(123456){\rm tr}(1234)-2 s_{12}s_{23}s_{34} {\rm tr}(4561) \bigr]\;I_5^n(p_{56},p_1,p_2,p_3,p_4)\nonumber\\
+\bigl[{\rm tr}(123456){\rm tr}(2345)-2 s_{23}s_{34}s_{45} {\rm tr}(5612) \bigr]\; I_5^n(p_{61},p_2,p_3,p_4,p_5) \bigr\} \nonumber
\end{eqnarray}
The coefficients $b_j$, $j \in \{1,\dots ,6\}$ are defined by the linear equation
\begin{eqnarray}\label{hexa_coeffs}
( \hat S \cdot b)_j = 1  &\Leftrightarrow& b_j = 
\sum\limits_{k=1}^6 \hat S^{-1}_{kj}\;\mbox{ where }\;\hat S_{kj}=(r_k-r_j)^2\\
\hat S &=& 
\left(\begin{array}{rrrrrr}
0      &  0      &  s_{23}  & s_{234}  &  s_{61}   &  0      \\
0      &  0      &  0       & s_{34}   &  s_{345}  &  s_{12} \\
s_{23} &  0      &  0       & 0        &  s_{45}   &  s_{123}\\ 
s_{234}& s_{34}  &  0       & 0        &  0        &  s_{56} \\
s_{61} &  s_{345}&  s_{45}  & 0        &  0        &  0      \\
0      &  s_{12}&  s_{123} &  s_{56}   &  0        &  0   
\end{array}\right) \nonumber\\
\det(\hat S)&=& 4 s_{12}s_{23}s_{34}s_{45}s_{56}s_{61} - {\rm tr}(123456)^2\nonumber
\end{eqnarray}
The traces allow for a compact notation for the coefficients $b_j$.
For completeness we also list the reduction formulas 
of pentagon and box integrals relevant for our
calculation in the appendix.
The  Gram matrix
$G_{kl} = 2 \,r_l \cdot r_k$ is related to 
$\hat S$ by $\hat S_{kl}=-G_{kl}+r_k^2+r_l^2$. 
For $N\ge 6$ and 4-dimensional external momenta one has det$(G)=0$, 
which leads to a non-linear constraint
between the Mandelstam variables. We note that this
constraint is represented {\em linearly} in terms of the
coefficients $b_j$. One has
\begin{eqnarray}\label{hexa_cons}
\det(G) = 0   \hspace{1cm} \Leftrightarrow  
\hspace{1cm} \sum\limits_{j=1}^6 b_j = 0
\end{eqnarray} 
By solving eq.~(\ref{hexa_coeffs}) with Cramer's rule
one sees that the constraint (\ref{hexa_cons}) relates
sums of determinants of 5 by 5 matrices. Expressing it in
terms of Mandelstam variables leads to a huge expression 
just representing zero.
The guideline to keep the sizes of expressions under control
in calculations of multi--particle processes
is thus to use representations of amplitudes where the $b_j$
are kept manifestly and to use relations (\ref{hexa_coeffs}) and 
(\ref{hexa_cons}) to perform cancellations as far as possible.

Applying the reduction formula (\ref{hexa_red}) above to reduce the 
hexagon, we observe that the coefficients of the hexagon and pentagon 
integrals in the amplitude combine in a nice way to form a 
resulting coefficient for a given pentagon which is again 
proportional to $b_j$. The resulting coefficient of 
$I_5^n(p_{12},p_3,p_4,p_5,p_6)$ in (\ref{amp_si}) is 
\begin{eqnarray}
{\rm tr}(3456) + {\rm tr}(123456) \, b_1 = -2 s_{34}s_{45}s_{56} \, b_4\;,
\end{eqnarray}
analogous for all cyclic permutations. 
Not only the size of the coefficients did not increase
but also one can still use linear relations. This will turn out
to be of major importance in what follows.
Now we reduce the pentagons to boxes 
using the reduction formula (\ref{red5}) given in the appendix. 
We obtain
\begin{eqnarray}
{\cal A}(p_1,p_2,p_3,p_4,p_5,p_6) =
  \frac{2}{3} \, I_3^n(p_{12},p_{34},p_{56})  \hspace{6cm}\nonumber\\
 + s_{12}\, I_4^n(p_1,p_2,p_{34},p_{56}) 
+\frac{{\rm tr}(14)}{4}I_4^n(p_1,p_{23},p_{4},p_{56})\nonumber\\
+\frac{b_2}{2E_2}\Big\{
s_{23}s_{34}\,[\,{\rm tr}(1234)-2s_{12}\,(s_{234}-s_{23})]\,
I_4^n(p_2,p_{3},p_{4},p_{561})\nonumber\\
+s_{12}s_{23}\,[\,{\rm tr}(1234)-2s_{34}\,(s_{123}-s_{23})]\,
I_4^n(p_1,p_{2},p_{3},p_{456})\nonumber\\
+{\rm tr}(1234)\,E_2\,I_4^n(p_1,p_{23},p_{4},p_{56})\nonumber\\
+s_{34}\,[-s_{123}\,{\rm tr}(1234)-2s_{12}s_{23}\,(s_{123}-s_{56})]\,
I_4^n(p_{3},p_{4},p_{56},p_{12})\nonumber\\
+s_{12}\,[-s_{234}\,{\rm tr}(1234)-2s_{23}s_{34}\,(s_{234}-s_{56})]\,
I_4^n(p_1,p_2,p_{34},p_{56})\Big\}\nonumber\\
+\quad 5 \mbox{ cyclic permutations}
\label{amp_box}
\end{eqnarray}
As a shorthand notation we use $E_1=s_{123}s_{345}-s_{12}s_{45}$. The $E_j$ 
for $j>1$ are defined by cyclic permutation.
Note that $E_{j}=E_{j+3}$.

The amplitude is now expressed in terms of four functions: The triangle 
with all three legs off-shell, box integrals with two off-shell legs at 
adjacent corners ($I_4^n(p_1,p_2,p_{34},p_{56})$ and 5 permutations), 
box integrals with two  off-shell legs at 
opposite corners ($I_4^n(p_1,p_{23},p_{4},p_{56})$ and 2 permutations), 
and box integrals with one off-shell leg 
($I_4^n(p_1,p_{2},p_{3},p_{456})$ and 5 permutations). 
We  now collect and combine the coefficients of particular terms in the 
cyclic sum. Already at this stage nontrivial cancellations
happen. First consider the coefficient $C^{op}$ of the "opposite" box 
$I_4^n(p_1,p_{23},p_{4},p_{56})=I_4^n(p_4,p_{56},p_{1},p_{23})$ 
in (\ref{amp_box}). It is given by 
\begin{eqnarray}
C^{op}&=&\frac{1}{2}\left\{{\rm tr}(14)+b_2\,{\rm tr}(1234)+
b_5\,{\rm tr}(4561)\right\}\nonumber\\
&=& (s_{56} + s_{23} -s_{234} -s_{123})  
- b_2 [ E_2 - s_{12} (s_{234}-s_{23}) - s_{34} (s_{123}-s_{23}) ]/2 \nonumber\\
&&- b_5 [ E_2 - s_{45} (s_{234}-s_{56}) - s_{61} (s_{123}-s_{56}) ]/2
\end{eqnarray}
Using $\hat S\cdot b=1$ to replace $b_2s_{12}$, $b_2s_{34}$, 
$b_5s_{45}$, $b_5 s_{61}$, one finds the useful relation
\begin{eqnarray}
2 \; C^{op} =  {\rm tr}(14)+b_2\,{\rm tr}(1234)+b_5\,{\rm tr}(4561)
= -E_2 \sum\limits_{j=1}^{6} b_j = 0\label{coop}
\end{eqnarray}
Hence the coefficients of the  box integrals with two  off-shell legs at 
opposite corners are identically zero!
One can combine the coefficients of the adjacent
boxes as a linear combination of $b_j$'s in a similar way.  
To investigate further cancellations we insert the expressions 
for the box integrals given in the appendix,
(\ref{1m}) and (\ref{ad}),  into (\ref{amp_box}) 
to obtain
\newpage
\begin{eqnarray}
{\cal A}(p_1,p_2,p_3,p_4,p_5,p_6) =
  \frac{2}{3} \, I_3^n(p_{12},p_{34},p_{56}) \hspace{7cm} \nonumber\\
 +\Big\{b_1-\frac{b_2}{2E_2}\,[\,{\rm tr}(1234)-2s_{34}\,(s_{123}-s_{23})]
-\frac{b_6}{2E_6}\,[\,{\rm tr}(5612)-2s_{56}\,(s_{345}-s_{61})]\Big\}\nonumber\\
\times\Big\{\frac{r_\Gamma}{\epsilon^2}
\left[
    (-s_{12})^{-\epsilon} 
+ \Bigl(  (-s_{234})^{-\epsilon} - (-s_{34})^{-\epsilon} \Bigr)
+ \Bigl(  (-s_{234})^{-\epsilon} - (-s_{56})^{-\epsilon} \Bigr)
\right] \nonumber\\
 -2 \,F_{2A}(s_{12},s_{234},s_{34},s_{56})  \Big\}\nonumber\\
+\Big\{\frac{b_1}{2E_1}\,[\,{\rm tr}(6123)-2s_{61}\,(s_{123}-s_{12})]
+ \frac{b_2}{2E_2}\,[\,{\rm tr}(1234)-2s_{34}\,(s_{123}-s_{23})] \Big\}\nonumber\\
\times \Big\{\frac{r_\Gamma}{\epsilon^2}
\left[
  (-s_{12})^{-\epsilon} 
+ (-s_{23})^{-\epsilon} 
+ \Bigl(  (-s_{12})^{-\epsilon} - (-s_{123})^{-\epsilon} \Bigr)  
+ \Bigl(  (-s_{23})^{-\epsilon} - (-s_{123})^{-\epsilon} \Bigr)
\right] \nonumber\\
 - 2 \, F_1(s_{12},s_{23},s_{123})   \Big\} 
+\quad 5 \mbox{ cyclic permutations}\hspace*{4.5cm} 
\label{amp_ex}
\end{eqnarray} 
The groupings of the pole terms are induced by the reduction formulas 
of box integrals, see appendix. Expression (\ref{amp_ex}) contains spurious double and single
poles. To see the cancellation of the pole terms it is enough to look at one 
double pole term, e.g. $(-s_{12})^{-\epsilon}/\epsilon^2$,
and one single pole term, e.g. 
$[(-s_{12})^{-\epsilon}-(-s_{123})^{-\epsilon}]/\epsilon^2$, separately.
The cancellation of the others then follows by cyclic symmetry.
The coefficient of $(-s_{12})^{-\epsilon}/\epsilon^2$ in the cyclic sum 
in (\ref{amp_ex}) is given by
\begin{eqnarray}
b_1+\frac{b_1}{2E_1}\left\{2\,{\rm tr}(1234)-2s_{61}\,(s_{123}-s_{12})-
2s_{23}\,(s_{345}-s_{12}) \right\}
=b_1+\frac{b_1}{2E_1}\left\{-2E_1\right\}=0
\end{eqnarray}
The coefficient of $[(-s_{12})^{-\epsilon}-(-s_{123})^{-\epsilon}]/\epsilon^2$ 
in the cyclic sum  (\ref{amp_ex}) is given by
\begin{eqnarray}
\frac{b_1}{2E_1}\left[\,{\rm tr}(6123)-2s_{61}\,(s_{123}-s_{12})\right]
+\frac{b_2}{2E_2}\left[2\,{\rm tr}(1234)-2s_{34}\,(s_{123}-s_{23})-
2s_{12}\,(s_{234}-s_{23})\right]\nonumber\\
+\frac{b_4}{2E_4}\left[\,{\rm tr}(3456)-2s_{56}\,(s_{345}-s_{45})\right]-
b_3
=-b_3+\frac{b_2}{E_2}(-E_2)-\frac{b_2+b_3}{E_1}(-E_1)=0
\label{cosingle}
\end{eqnarray}
where again $\hat S\cdot b=1$ and (\ref{coop}) have been used. 
The remaining structures are now $I_3^n$, $F_{2A}$ and $F_1$.
The latter two contain dilogarithms with single ratios
of Mandelstam variables, products of logarithms and $\pi^2$
terms. The fact that the single poles stemming from the differences 
$[(-s_{12})^{-\epsilon}-(-s_{123})^{-\epsilon}]/\epsilon^2$ 
(which in turn are related to the triangles with two off-shell 
external legs occurring in the reduction of box integrals) cancel 
independently from those stemming from the double pole terms (like 
$(-s_{12})^{-\epsilon}/\epsilon^2$, which are related to the triangles 
with one off-shell external leg) has an important consequence for the 
finite part of the amplitude: By examination of expressions (\ref{1m}) 
to (\ref{f2a}) for the box integrals one observes that a term 
$[(-a)^{-\epsilon}-(-b)^{-\epsilon}]/\epsilon^2$ always is 
associated with a dilogarithm  
$Li_2(1-a/b)$ in the finite part of the same box integral. 
Therefore the cancellation of the terms  
$[(-a)^{-\epsilon}-(-b)^{-\epsilon}]/\epsilon^2$ in the amplitude 
immediately leads to the cancellation of the dilogarithms. 
The $\pi^2$ terms present in the box with one off-shell leg also cancel 
in  (\ref{amp_ex}) due to relations (\ref{hexa_coeffs}) and (\ref{coop}). 
Hence the only terms which survive are the triangle graphs and
some logarithmic terms stemming from the finite parts of the 
box integrals, such that we finally obtain 
\newpage
\begin{eqnarray}\label{amp_final}
{\cal A}(p_1,p_2,p_3,p_4,p_5,p_6) = G(p_1,p_2,p_3,p_4,p_5,p_6)
+ 5 \; \mbox{cyclic permutations}
\end{eqnarray}
with
\begin{eqnarray} G(p_1,p_2,p_3,p_4,p_5,p_6) = 
 \frac{2}{3} \,I^n_3(p_{12},p_{34},p_{56})\hspace{6cm} \nonumber \\
+ \left\{  \frac{b_1}{E_1} 
\left[{\rm tr}(6123)-2s_{61}(s_{123}-s_{12}) \right] 
+\frac{b_2}{E_2} 
\left[ {\rm tr}(1234)-2s_{34}(s_{123}-s_{23}) \right] \right\} 
\hspace{1cm}\nonumber \\ \times
\log\left( \frac{s_{12}}{s_{123}} \right)
         \log\left( \frac{s_{23}}{s_{123}} \right) \nonumber \\
+\left\{ -b_1 + \frac{b_2}{2 E_2} 
\left[ {\rm tr}(1234) - 2 s_{34} (s_{123}-s_{23})  \right]
+ \frac{b_6}{2 E_6}
\left[ {\rm tr}(5612)- 2 s_{56} (s_{345}-s_{61})\right]\right\}
\hspace{1cm}\nonumber \\ \times
\left[ \log\left( \frac{s_{12}}{s_{234}} \right) 
       \log\left( \frac{s_{56}}{s_{234}} \right)
     + \log\left( \frac{s_{34}}{s_{234}} \right) 
       \log\left( \frac{s_{12}}{s_{56}} \right) \right]\nonumber\\
       \label{ampsi}
\end{eqnarray}
Note that $G(p_1,p_2,p_3,p_4,p_5,p_6)$ has no spurious singularities. 
We checked that the numerator of expression (\ref{ampsi}) vanishes in 
the limits where its denominator  vanishes. 

Finally, the full amplitude is given by the sum over
permutations of the function $G$
\begin{eqnarray}\label{Eq:final}
\Gamma_{\rm yuk}^{\phi}[p_1,p_2,p_3,p_4,p_5,p_6] 
= - \frac{g^6}{(4\pi)^2}\sum\limits_{\pi\in S_6}^{} G(p_{\pi_1},p_{\pi_2},
                  p_{\pi_3},p_{\pi_4},p_{\pi_5},p_{\pi_6})
\end{eqnarray} 
From (\ref{amp_graph}) it is clear that
half of the $6!$ permutations simply correspond to a  change in orientation
of the fermion line which does not change the value of the integral. 
It is thus enough to sum in (\ref{Eq:final}) over orientation conserving permutations, i.e. $\pi\in S_6/Z_2$, and to multiply by a factor two.

\section{Conclusion}

We calculated a certain class of hexagon amplitudes 
in the Yukawa model. First, $N$--point amplitudes 
with scalars/pseudoscalars as external particles attached to 
a fermion loop were constructed using string inspired methods.
The amplitudes turned out to be represented in terms
of scalar integrals only. 
Thus the Yukawa model is an adequate testing ground to study 
nontrivial cancellations appearing in scalar integral reductions
in isolation from additional complications due to a nontrivial
tensor structure in more realistic situations such as gauge theory 
amplitudes. Focusing on the massless case and $N=6$ for a 
representative amplitude we first 
demonstrated the equivalence of the string inspired to the 
Feynman diagrammatic approach, then we explicitly calculated
the amplitude. This was done by using reduction formulas for scalar
$N$--point integrals.
It was shown in detail how cancellations can be made manifest
at each step of the calculation by using linear relations
between reduction coefficients. This saved us from dealing
with large expressions at any stage of the calculation.
With the present method there is no explosion of terms typical 
for multi--leg  calculations. The final answer is surprisingly compact
and contains --- apart from 3--point functions with 
3 off--shell legs --- only some products of logarithms.
A reason for that lies certainly in the fact that the amplitudes
under consideration are infrared finite.

In the case of off--shell amplitudes 
the increasing number of kinematic invariants will lead
to larger expressions. Still, reduction coefficients
will obey linear relations similar to the ones used in deriving
the on--shell amplitudes.
The same is true in the case of massive particles.
Thus one can expect analogous cancellation mechanisms in   
both cases. This deserves further study.  

As a next step more realistic examples have to be considered
including gauge bosons and a nontrivial infrared structure. 
Again, it is justified to speculate that
the recombination of scalar integrals
will work similarly. Hopefully this work is a 
step towards efficient algorithms to calculate 
multi--particle amplitudes at one loop.  

\section*{Acknowledgements}
This work was supported in part by the EU Fourth Training Programme  
''Training and Mobility of Researchers'', Network ''Quantum Chromodynamics
and the Deep Structure of Elementary Particles'',
contract FMRX--CT98--0194 (DG 12 - MIHT) and also by
CONACyT (Mexico) and CNRS (France), grant number CONACyT E130.917/2001.  

\section*{Appendix}
\subsection*{Explicit reduction formulas for scalar integrals}
We collect here reduction formulas relevant for the computation of
massless hexagon amplitudes. The given formulas are sufficient to 
deal with any scalar integral arising in the calculation of massless hexagon amplitudes.

The pentagon integrals with one external leg off-shell are reduced 
by the formula
\begin{eqnarray}
I_5^n(p_{12},p_3,p_4,p_5,p_6) \quad = \qquad\frac{1}{2E_1}\Big\{
\frac{1}{s_{34}}\,[-{\rm tr}(3456)+2s_{34}(s_{123}-s_{45})]\,
I_4^n(p_4,p_5,p_6,p_{123})\nonumber\\
+\frac{1}{s_{56}}\,[-{\rm tr}(3456)+2s_{56}(s_{345}-s_{45})]\,
I_4^n(p_3,p_4,p_5,p_{612})\nonumber\\
+\frac{1}{s_{34}s_{45}s_{56}}\,[-{\rm tr}(3456)\,E_1]\,
I_4^n(p_6,p_{12},p_3,p_{45})\nonumber\\
+\frac{1}{s_{34}s_{45}}\,[s_{345}\,{\rm tr}(3456)+2s_{34}s_{45}(s_{345}-s_{12})]\,
I_4^n(p_5,p_6,p_{12},p_{34}) \nonumber\\
+\frac{1}{s_{45}s_{56}}\,[s_{123}\,{\rm tr}(3456)+2s_{45}s_{56}(s_{123}-s_{12})]\,
I_4^n(p_3,p_4,p_{56},p_{12}) \Big\}\label{red5}
\end{eqnarray}
where $E_1 = s_{123}s_{345}-s_{12}s_{45}$.
For the boxes three cases have to be distinguished: 2 off-shell
legs at opposite corners, 2 off--shell
legs at adjacent corners, and 1 off--shell leg. 
Note that the infrared poles of the boxes are contained in 
 triangle graphs with one and/or two legs off--shell.
The off--shell momenta are  sums of light--like vectors.
For the "adjacent" case we find
\begin{eqnarray} 
I_4^n(p_1,p_2,p_{34},p_{56}) = 
\frac{2s_{34}s_{56}+s_{234} ( s_{12}-s_{56}-s_{34})}{s_{234}^2 s_{12}}
I_3^n(p_{12},p_{34},p_{56})\nonumber\\
+\frac{s_{234}-s_{56}}{s_{234}s_{12}} I_3^n(p_1,p_{234},p_{56}) 
+\frac{1}{s_{234}} I_3^n(p_1,p_2,p_{3456}) 
+\frac{s_{234}-s_{34}}{s_{234}s_{12}} I_3^n(p_2,p_{34},p_{561}) \nonumber\\
+2(n-3)\frac{s_{34}s_{56}-s_{234}(s_{34}+s_{56}-s_{12}-s_{234})}{s_{12}s_{234}^2} I_4^{n+2}(p_1,p_2,p_{34},p_{56}) \nonumber\\
\end{eqnarray}
\newpage
The "opposite" case gives 
\begin{eqnarray} 
I_4^n(p_1,p_{23},p_4,p_{56}) = \hspace{8cm} \nonumber\\
\quad\frac{s_{123}-s_{56}}{s_{123}s_{234}-s_{23}s_{56}} I_3^n(p_4,p_{56},p_{123})
+\frac{s_{234}-s_{56}}{s_{123}s_{234}-s_{23}s_{56}} I_3^n(p_1,p_{234},p_{56})\nonumber\\
+\frac{s_{123}-s_{23}}{s_{123}s_{234}-s_{23}s_{56}} I_3^n(p_1,p_{23},p_{456})
+\frac{s_{234}-s_{23}}{s_{123}s_{234}-s_{23}s_{56}} I_3^n(p_4,p_{561},p_{23})\nonumber\\
+2(n-3)\frac{s_{234}+s_{123}-s_{23}-s_{56}}{s_{123}s_{234}-s_{23}s_{56}}
I_4^{n+2}(p_1,p_{23},p_4,p_{56}) \nonumber\\
\end{eqnarray}
and finally the case with one leg off--shell
\begin{eqnarray} 
I_4^n(p_1,p_2,p_3,p_{456}) = 
\quad\frac{s_{12}-s_{123}}{s_{12}s_{23}} I_3^n(p_3,p_{456},p_{12})
+\frac{s_{23}-s_{123}}{s_{12}s_{23}} I_3^n(p_1,p_{23},p_{456})\nonumber\\
+\frac{1}{s_{23}} I_3^n(p_1,p_2,p_{3456})
+\frac{1}{s_{12}} I_3^n(p_2,p_3,p_{4561})\nonumber\\
+2(n-3)\frac{s_{12}+s_{23}-s_{123}}{s_{12}s_{23}}
I_4^{n+2}(p_1,p_2,p_3,p_{456}) \nonumber\\
\end{eqnarray}
All dilogarithms are collected in the terms $I_4^{n+2}$ and the 
triangles with 3 legs off-shell. In the case of the box with two adjacent
legs off--shell, $I_3^n$ and $I_4^{n+2}$ combine to a much simpler
expression than the single expressions individually. 
This indicates that the splitting into triangles and remainder terms
is only useful, if infrared singularities are present.
Explicit formulas for the scalar integrals are given below. 

\subsection*{List of scalar integrals}
The triangles with 
one and two on--shell legs are given by
\begin{eqnarray}
I_3^n(p_1,p_2,p_{3456}) &=&  \frac{r_\Gamma}{\epsilon^2}  \frac{(-s_{12})^{-\epsilon}}{s_{12}} \\
I_3^n(p_1,p_{23},p_{456}) &=&  \frac{r_\Gamma}{\epsilon^2}  \frac{(-s_{23})^{-\epsilon}-(-s_{123})^{-\epsilon}}{s_{23}-s_{123}}   \\
r_\Gamma &=& \frac{\Gamma(1+\epsilon)\Gamma(1-\epsilon)^2}{\Gamma(1-2\epsilon)}\nonumber
\end{eqnarray}
For the finite triangle  
with all legs off--shell we quote
the four dimensional representation~\cite{Ghinculov}
\begin{eqnarray}
I_3(p_{12},p_{34},p_{56}) &=& -\frac{1}{\sqrt{\lambda}}
\left\{
2 \,Li_2\left(1-\frac{1}{y_2}\right) + 2 \,Li_2\left(1-\frac{1}{x_2}\right) +
\frac{\pi^2}{3} \right.
\nonumber\\
&& \left. +\frac{1}{2}
\left[ \log^2\left( \frac{x_1}{y_1}\right)
      +\log^2\left( \frac{x_2}{y_2}\right)
      -\log^2\left( \frac{x_2}{y_1}\right)
      +\log^2\left( \frac{x_1}{y_2}\right) \right]
\right\} \label{tri}\\
x_{1,2} &=& \frac{s_{12}+s_{34}-s_{56}\mp\sqrt{\lambda}}{2 s_{12}}\nonumber\\
y_{1,2} &=& \frac{s_{12}-s_{34}+s_{56}\pm\sqrt{\lambda}}{2 s_{12}}\nonumber\\ 
\lambda &=& s_{12}^2 + s_{34}^2 + s_{56}^2 -2 s_{12}s_{34}-
2 s_{34}s_{56}-2 s_{56}s_{12} - i \,\mbox{sign}(s_{12})\delta\nonumber
\end{eqnarray}
The  infinitesimal imaginary assures that 
the formula is valid in all kinematic regions by using
\begin{eqnarray}
\sqrt{\lambda \pm i\delta } = \left\{ 
\begin{array}{cc}
\sqrt{\lambda} \pm i\delta  & , \; \lambda \geq 0 \\
\pm\, i\, \sqrt{-\lambda}   & , \; \lambda < 0
\end{array}
\right.
\end{eqnarray}
In the splitting of the box integrals into divergent and finite pieces
the grouping of the $(-s_{ij})^{-\epsilon}$ terms  is induced by the 
triangle graphs. We keep this form in which the 
separation of single and double poles is manifest.
\begin{eqnarray} 
I_4^n(p_1,p_2,p_3,p_{456}) = \frac{1}{s_{12}s_{23}} 
\left\{ 
\frac{r_\Gamma}{\epsilon^2}
\left[
  (-s_{12})^{-\epsilon} 
+ (-s_{23})^{-\epsilon} \right.\right. \hspace{4cm}\nonumber\\
\left.\left.
+ \Bigl(  (-s_{12})^{-\epsilon} - (-s_{123})^{-\epsilon} \Bigr)  
+ \Bigl(  (-s_{23})^{-\epsilon} - (-s_{123})^{-\epsilon} \Bigr)
\right] 
- 2 \, F_1(s_{12},s_{23},s_{123})  \right\} \label{1m}
\end{eqnarray}
\begin{eqnarray} 
I_4^n(p_1,p_2,p_{34},p_{56}) = \frac{1}{s_{12}s_{234}} \hspace{8cm} \nonumber\\
\Big\{ 
\frac{r_\Gamma}{\epsilon^2}
\left[
    (-s_{12})^{-\epsilon} 
+ \Bigl(  (-s_{234})^{-\epsilon} - (-s_{34})^{-\epsilon} \Bigr)
+ \Bigl(  (-s_{234})^{-\epsilon} - (-s_{56})^{-\epsilon} \Bigr)
\right] \nonumber\\
 -2 \,F_{2A}(s_{12},s_{234},s_{34},s_{56}) \Big\}\label{ad}
\end{eqnarray}
\begin{eqnarray} 
I_4^n(p_1,p_{23},p_4,p_{56}) = \frac{1}{s_{123}s_{234}-s_{23}s_{56}} \hspace{6cm} \nonumber\\
\Big\{ 
\frac{r_\Gamma}{\epsilon^2}
\left[
  \Bigl(  (-s_{123})^{-\epsilon} - (-s_{23})^{-\epsilon} \Bigr)
+ \Bigl(  (-s_{123})^{-\epsilon} - (-s_{56})^{-\epsilon} \Bigr) 
\right. \hspace{2cm}\nonumber\\ \left.
+ \Bigl(  (-s_{234})^{-\epsilon} - (-s_{23})^{-\epsilon} \Bigr)
+ \Bigl(  (-s_{234})^{-\epsilon} - (-s_{56})^{-\epsilon} \Bigr)
\right]\nonumber\\
 -2\,F_{2B}(s_{123},s_{234},s_{23},s_{56}) \Big\}\label{op}
\end{eqnarray}
The finite terms are given by  logarithms and dilogarithms.
\begin{eqnarray} 
F_1(s_{12},s_{23},s_{123}) =  -\,Li_2\left( 1-\frac{s_{12}}{s_{123}}\right)
- \,Li_2\left( 1-\frac{s_{23}}{s_{123}}\right) \nonumber \\
-\log\left( \frac{s_{12}}{s_{123}}\right)\log\left( \frac{s_{23}}{s_{123}}\right)
+\frac{\pi^2}{6}
\label{f1} 
\end{eqnarray}
\begin{eqnarray}
F_{2A}(s_{12},s_{234},s_{34},s_{56}) =
\,Li_2\left( 1-\frac{s_{34}}{s_{234}}\right)
+\,Li_2\left( 1-\frac{s_{56}}{s_{234}}\right)\hspace{2.5cm}\nonumber\\
+\frac{1}{2}\log\left( \frac{s_{12}}{s_{234}}\right) 
            \log\left( \frac{s_{56}}{s_{234}}\right)
+\frac{1}{2}\log\left( \frac{s_{34}}{s_{234}}\right) 
            \log\left( \frac{s_{12}}{s_{56}}\right)	    
\label{f2a}
\end{eqnarray}
\begin{eqnarray}
F_{2B}(s_{123},s_{234},s_{23},s_{56}) =
- \,Li_2\left( 1-\frac{s_{23}s_{56}}{s_{123}s_{234}}\right)
+ \,Li_2\left( 1-\frac{s_{23}}{s_{123}}\right)\hspace{1.5cm}\nonumber\\
+ \,Li_2\left( 1-\frac{s_{23}}{s_{234}}\right)
+ \,Li_2\left( 1-\frac{s_{56}}{s_{123}}\right)
+ \,Li_2\left( 1-\frac{s_{56}}{s_{234}}\right)
+\frac{1}{2}\log^2\left( \frac{s_{123}}{s_{234}}\right)
\label{f2b}
\end{eqnarray}
\newpage
The pentagon integral is now expressible as
a pole part and a remainder as follows:
\begin{eqnarray}
I_5^n(p_{12},p_2,p_3,p_4,p_5,p_6) = 
I_5^n(p_{12},p_2,p_3,p_4,p_5,p_6)\vert_{\mathrm{pole \,part}} \hspace{2cm}\nonumber\\
- \frac{1}{s_{34}s_{45}s_{56}} 
  F^{Penta}_{1}(s_{123},s_{34},s_{45},s_{56},s_{345},s_{12})
\end{eqnarray}
\begin{eqnarray} 
I_5^n(p_{12},p_2,p_3,p_4,p_5,p_6)\vert_{\mathrm{pole \,part}} = \hspace{7cm}\nonumber\\
\frac{r_\Gamma}{\epsilon^2} 
\Bigl\{
 \frac{(-s_{34})^{-\epsilon}}{s_{123}s_{34}s_{45}}
  +\frac{(-s_{45})^{-\epsilon}}{s_{34}s_{45}s_{56}} 
  +\frac{(-s_{56})^{-\epsilon}}{s_{45}s_{56}s_{345}} \hspace{5cm}\nonumber\\
+\frac{(s_{56}-s_{123})[(-s_{123})^{-\epsilon}-(-s_{56})^{-\epsilon}]}{s_{34}s_{45}s_{56}s_{123} }
  +\frac{(s_{34}-s_{345})[(-s_{345})^{-\epsilon}-(-s_{34})^{-\epsilon}]}{s_{34}s_{45}s_{56}s_{345} } 
\hspace{0.2cm}\nonumber\\
+\frac{(s_{123}-s_{12})[(-s_{123})^{-\epsilon}-(-s_{12})^{-\epsilon}]}{s_{34}s_{123} ( s_{123}s_{345}-s_{12}s_{45} )}
  +\frac{(s_{345}-s_{12})[(-s_{345})^{-\epsilon}-(-s_{12})^{-\epsilon}]}{s_{56}s_{345} ( s_{123}s_{345}-s_{12}s_{45} )} 
\hspace{0.2cm}\nonumber\\ 
+\frac{(s_{123}-s_{45})[(-s_{45})^{-\epsilon}-(-s_{123})^{-\epsilon}]}{s_{45}s_{56}  ( s_{123}s_{345}-s_{12}s_{45} )}
  +\frac{(s_{345}-s_{45})[(-s_{45})^{-\epsilon}-(-s_{345})^{-\epsilon}]}{s_{34}s_{45}  ( s_{123}s_{345}-s_{12}s_{45} )} 
\Bigr\} \label{pentapole}
\end{eqnarray}
\begin{eqnarray}
F^{Penta}_{1}(s_{123},s_{34},s_{45},s_{56},s_{345},s_{12}) = \hspace{6cm} \nonumber\\ \frac{1}{E_1}
\Bigl\{
 \left[ E_1 + s_{34}(s_{123}-s_{45}) -s_{56}(s_{345}-s_{45})  \right] \, 
F_1(s_{45},s_{56},s_{123})  \nonumber\\
+ \left[ \frac{(s_{34} s_{45})}{s_{345}}(s_{345}-s_{12})-\frac{(s_{345}-s_{34})}{s_{345}}E_1 +s_{56}(s_{345}-s_{45})\right] \, 
F_{2A}(s_{56},s_{345},s_{12},s_{34})\nonumber\\
+\left[ E_1 -s_{34}(s_{123}-s_{45}) -s_{56}(s_{345}-s_{45})\right] \, F_{2B}(s_{345},s_{123},s_{12},s_{45})\nonumber\\
+\left[ \frac{(s_{45} s_{56})}{s_{123}}(s_{123}-s_{12})-\frac{(s_{123}-s_{56})}{s_{123}}E_1 +s_{34}(s_{123}-s_{45}) \right] \, 
F_{2A}(s_{34},s_{123},s_{56},s_{12})\nonumber\\ 
+\left[ E_1- s_{34}(s_{123}-s_{45})  +s_{56}(s_{345}-s_{45})  \right] \, F_{1}(s_{34},s_{45},s_{345}) \Bigr\}\nonumber\\
\end{eqnarray}

\end{document}